\documentclass[runningheads,a4paper]{llncs}

\usepackage{amssymb}
\usepackage{graphicx}

\usepackage{algorithm}
\usepackage{algorithmicx}
\usepackage{algpseudocode}
\usepackage{color}
\usepackage{wrapfig}
\usepackage{multicol}

\usepackage{url}
\urldef{\mailsa}\path|salboaie@gmail.com|
\urldef{\mailsb}\path|doina.cosovan@info.uaic.ro|
\newcommand{\keywords}[1]{\par\addvspace\baselineskip
	\noindent\keywordname\enspace\ignorespaces#1}
\newcommand{\squeezeup}{\vspace{-3mm}}

\begin{document}
\mainmatter
\title{Private Data System Enabling Self-Sovereign\thanks{Author's version. The final publication is available at Springer via http://dx.doi.org/10.1007/978-3-319-59665-5\_6.}\\Storage Managed by Executable Choreographies}
\titlerunning{Private Data System}
\author{Sinic\u{a} Alboaie$^{1,2}$ \and Doina Cosovan$^1$}
\authorrunning{Sinic\u{a} Alboaie, Doina Cosovan}
\institute{$^1$Alexandru Ioan Cuza University of Iasi, Romania\\
	$^2$Technical University of Cluj-Napoca, Romania\\
	\mailsa \hspace{5pt} \mailsb\\
}
\maketitle
\squeezeup
\squeezeup
\squeezeup
\begin{abstract}
With the increased use of Internet, governments and large companies store and share massive amounts of personal data in such a way that leaves no space for transparency. When a user needs to achieve a simple task like applying for college or a driving license, he needs to visit a lot of institutions and organizations, thus leaving a lot of private data in many places. The same happens when using the Internet. These privacy issues raised by the centralized architectures along with the recent developments in the area of serverless applications demand a decentralized private data layer under user control.

We introduce the Private Data System (PDS), a distributed approach which enables self-sovereign storage and sharing of private data. The system is composed of nodes spread across the entire Internet managing local key-value databases. The communication between nodes is achieved through executable choreographies, which are capable of preventing information leakage when executing across different organizations with different regulations in place.

The user has full control over his private data and is able to share and revoke access to organizations at any time. Even more, the updates are propagated instantly to all the parties which have access to the data thanks to the system design. Specifically, the processing organizations may retrieve and process the shared information, but are not allowed under any circumstances to store it on long term.

PDS offers an alternative to systems that aim to ensure self-sovereignty of specific types of data through blockchain inspired techniques but face various problems, such as low performance. Both approaches propose a distributed database, but with different characteristics. While the blockchain-based systems are built to solve consensus problems, PDS's purpose is to solve the self-sovereignty aspects raised by the privacy laws, rules and principles.

\squeezeup
\end{abstract}

\squeezeup
\squeezeup
\squeezeup
\squeezeup
\section{Introduction}
\squeezeup

Every time a user needs to create an account, he needs to provide a lot of private information, like name, birth date, gender, marital status, and so on. Even more, he needs to choose and answer to some security questions for account recovery in case he forgets the password or simply for user validation when performing sensitive actions. These security questions usually consist of private data as well.

This way, each user spreads his private data to a lot of organizations / companies / service providers. This raises two main issues: one is related to data protection and the other - to data duplication. In regards to data protection, each organization has its own ways of storing and protecting the data. Some are better than others. The user's data is as safe as the weakest organization to which the user provided his data. Thus an attacker can target the weakest link to learn private information. The data duplication issue consists mainly of the fact that changing a piece of private information (like changing the last name by getting married) requires updating it in all the places this particular piece of private information was saved, which is burdening and time consuming.

An existing way of solving these problems is by using single sign-on techniques. But this makes the user dependent on the single sign-on provider because losing access to the account used for single sign-on means losing access to all the accounts authenticating the user with this single sign-on account.

We propose a solution that enables users to keep full control over their private data. Private Data System (PDS) is a distributed scalable system composed of three types of nodes, transparently spread across the entire Internet: audit, index and storage nodes. Each node manages a local key-value database and each type of node has its own purpose in the system, as explained further.

Each piece of private information is split into undecipherable chunks. Each chunk is assigned a different partial key in a different key-value database managed by a different storage node. The association between these keys is stored under a master key in a key-value database managed by an index node. Since the data needs to be accessible by different processing nodes, the master key is referenced by different key references. The association between them is stored under a key reference in a key-value database managed by an audit node. The key references are the only points of access to the actual data. Hence it also contains information regarding who owns the referenced data, who was this particular key reference shared with, and metadata describing the referenced data.

The communication between nodes is achieved with the help of executable choreographies, which visit the needed nodes in the needed order, execute on each node the needed operations, and return to the user with the results.

An interesting use case of using PDS are social networks and other systems which, besides private data, manage also trust and reputation data \cite{alboaie2011trust}.

We will start by reviewing the related work (Section \ref{sec:related_work}) and introducing the system along with its elements (Section \ref{sec:system_elems}). Then, we'll explain how CRUD (create, read, update, delete) and sharing / revoking operations work (Section \ref{sec:system_ops}). In the end, we will analyze the proposed system from the privacy perspective (Section \ref{sec:system_analysis}), conclude (Section \ref{sec:conclusions}), and present future directions in regards to the proposed system (Section \ref{sec:future_work}).

\squeezeup
\section{Related Work}
\label{sec:related_work}
\squeezeup

Smart systems integrate technology, organizations and people in order to accomplish complex processes that are controlled by computer systems. For a large number of integration points, integration is achieved through classical ESB (Enterprise Service Bus) - type systems \cite{DBLP:books/daglib/0011231}, MOM (Message-Oriented Middleware) systems \cite{curry2004message}, systems based on EIP (Enterprise Integration Patterns) \cite{hohpe2004enterprise} or through the orchestration of services through custom code or languages used to model business processes \cite{DBLP:journals/crossroads/Ko09}.

All these methods tend to be sufficient to integrate the components belonging to one organization. On the other hand, the integration among multiple organizations should be addressed using choreographies as any centralized solution is risky in terms of security and private data protection. Composition of systems using orchestration tends to create centralized systems.

Although many authors perceive choreographies as a mechanism to describe in a more formal way the contracts among several organizations \cite{WSCDLSpec}, the academic research proposed the concept of executable choreographies \cite{akkawi2006executable} \cite{alboaie2015levels} \cite{besana2009executable}. They suggest transforming the descriptions of the choreographies in code that is executed inside each organization participating in the choreography. As such, a choreography is not only a formal description of a contract among organizations but also a description of a workflow in an executable way. The same description (choreography) gets to run in several organizations in a decentralized manner (without the need for a centralized conductor) and therefore any need to translate the choreography into other programming languages disappears.

While PDS could be implemented outside the world of the executable choreographies, we believe that choreographies are suitable for the complex workflows operating across multiple organizations. The code of the executable choreographies is verifiable at a higher level and can provide confidence that the implementation provides the privacy properties of the theoretical model.

Another advantage of the executable choreographies is that it comes with a solution for the self sovereign identity. The Sovrin Foundation explains in \cite{Sovrin} why the rise of the self sovereign identities was inevitable and details the path that had to be traversed for the community to come to this conclusion.

For PDS, the data owner must be identified and authorized. Supplementary benefits in the data leakage preventions could be achieved if the data owner is identified in all the other organizations contributing to a request without leaking its identity (using some anonymous aliases controlled by the data owner). The executable choreographies aim at offering these benefits without any supplementary implementation effort. However, detailing the way in which the self sovereign identities used by the data owners and processors are authenticated and authorized is not the purpose of this paper. It is a complex enough topic to require its own paper, so we will revisit this issue at a later date.

In regards to the advances related to data sovereignty, we would like to mention \cite{esposito2016encryption}, which proposes storing encrypted data in cloud federations and \cite{agrawal2003information}, which proposes sovereign information sharing in order to integrate the information belonging to autonomous entities. Queries are executed on the databases and reveal only the results. The work is continued in \cite{agrawal2004enabling} which enables sovereign information sharing using web services. This work applies to service providers which want to allow queries on their databases without sharing the content on which the queries are executed. Our work focuses on the average user which needs to own and store his data in a single place and provide / revoke access to it to various service providers as needed.

Note that \cite{peterson2011position} introduces the data sovereignty notion for establishing the nation-state where the cloud storage service providers are storing the data physically in order to ensure they are meeting their contractual geographic obligations. In this paper, we consider data sovereignty to be the ability of the user to have full control over his data and the entities to which it is shared or revoked.

States and international organizations start to gradually introduce principles and standards, the most notable being Privacy By Design \cite{EUDataProtection}. Collecting information in parallel with the absence of technical constraints on how companies can use the data intentionally or unintentionally begins to be perceived as a risk. On the one hand, there are risks for companies because users could refuse to adopt privacy challenged technologies. On the other hand, we have risks regarding the whole society, the most obvious being represented by the potential that some companies can influence society in illegal and immoral manners.

Commercial exploitation of private data has come to create the impression that people are exploited commercially in ways that do not adequately compensate for the risks they take. A more transparent model that allows fair and equitable use of personal data is needed. Considering all these aspects, the article proposes a software architecture in which private data's storage places are under the strict control of the user or his delegates.

\squeezeup
\section{System Elements}
\label{sec:system_elems}

In this section, we define the terminology used for the Private Data System throughout this paper. First, we define the following \textbf{roles}:\\
\textbf{Data Owner (DO)} represents the identity which owns the data.\\
\textbf{Data Processor (DP)} represents the identity which processes the data; the identity to which the data was shared.

Second, we define the following \textbf{types of data}:\\
\textbf{Private Data (PD)} represents the private data which is to be stored in the system; if a piece of private data PD is split into n undecipherable chunks, then PDi, i = 0,n is an undecipherable chunk of data.\\
\textbf{Metadata (MD)} specifies the relationship between the Private Data and Data Processors by labeling the data according to the Data Owner and ontologies.

Third, since the system is based on key-value databases, we define the following \textbf{keys for data storage, associations, and references}:\\
\textbf{Master Key (MK)} represents and anonymizes a piece of private information.\\
\textbf{Partial Key (PK)} represents and anonymizes one undecipherable chunk from the set of undecipherable chunks in which a piece of private information was split. Thus, the MK is associated to the set of PKs which represent the set of undecipherable chunks needed to recombine the piece of private information.\\
\textbf{Key Reference (KR)} represents a reference to / an alias of a piece of private information (a reference to a Master Key).\\
\textbf{Key Reference Hash (KRH)} is obtained by applying a hash function on the Key Reference value and adding the address of the processing node which is to receive the results.

In the end, we define the following types of nodes:\\
\textbf{Processing Node (PN)} stores Key References and needs to retrieve and process the private data referenced by them. Processing nodes are forbidden by law to store the retrieved data on long term.\\
\textbf{Audit Node (AN)} manages a key-value database which stores the association between Key References and the Master Keys they reference along with the information describing the data referenced by the Master Key (Data Owner, Data Processor, and Metadata). In the database, the key is a Key Reference and the value is a tuple consisting of the Master Key, the Metadata, the Data Owner, and the Data Processor.\\
\textbf{Index Node (IN)} manages a key-value database which stores the association between Master Keys and its corresponding Partial Keys. In the database, the key is the Master Key and the value is the list of Partial Keys needed to reconstruct the piece of private information represented by the Master Key.\\
\textbf{Storage Node (SN)} manages a key-value database which stores the association between Partial Keys and Partial Messages. In the database, the key is the Partial Key and the value is the undecipherable chunk of data represented by this particular Partial Key.

\section{System Operations}
\label{sec:system_ops}

In this section we detail the way in which CRUD (Create, Read, Update, Delete) operations as well as copying, sharing, and revoking access to data are performed in the proposed system. For simplicity, we are going to use the following notations throughout this paper:
\begin{itemize}
	\item $[E_1, E_2, ..., E_n]$ is a list containing the elements $E_1, E_2, ..., E_n$.
	\item $(E_1, E_2, ..., E_n)$ is a tuple containing the elements $E_1, E_2, ..., E_n$.
	\item $\{K_1: V_1, K_2: V_2, ..., K_n: V_n\}$ is a dictionary in which the value $V_1$ is stored under the key $K_1$, the value $V_2$ is stored under the key $K_2$, ..., and the value $V_n$ is stored under the key $K_n$.
	\item $N_1 \rightarrow N_2: M$ means the node $N_1$ sends to the node $N_2$ the message $M$, which corresponds to performing a step in the executable choreography.
	\item $DB[K] := V$ means the value $V$ is stored under the key $K$ in the key-value database $DB$ by the node managing $DB$.
	\item $V := DB[K]$ means the value $V$ associated to the key $K$ is retrieved from the key-value database $DB$ by the node managing $DB$.
	\item $N_1: A$ means the node $N_1$ performs the action $A$.
	\item $M := gen()$ means the message $M$ is generated (either randomly or according to an algorithm); this is an action.
	\item $PD_1, PD_2, ..., PD_n := split(PD)$ means the private data $PD$ is split into $n$ undecipherable chunks of data $PD_1, PD_2, ..., PD_n$; this is an action.
	\item $PD := recombine(PD_1, PD_2, ..., PD_n)$ means the $n$ undecipherable chunks of data $PD_1, PD_2, ..., PD_n$ are recombined in order to obtain the initial piece of private data $PD$ which was split to obtain them; this is an action.
\end{itemize}
\squeezeup
\squeezeup
\subsection{Creating / Storing Private Data}
\begin{figure}[htbp]
	\begin{center}
		\caption{Storing Private Data}\centering
		\includegraphics[width=4.5in]{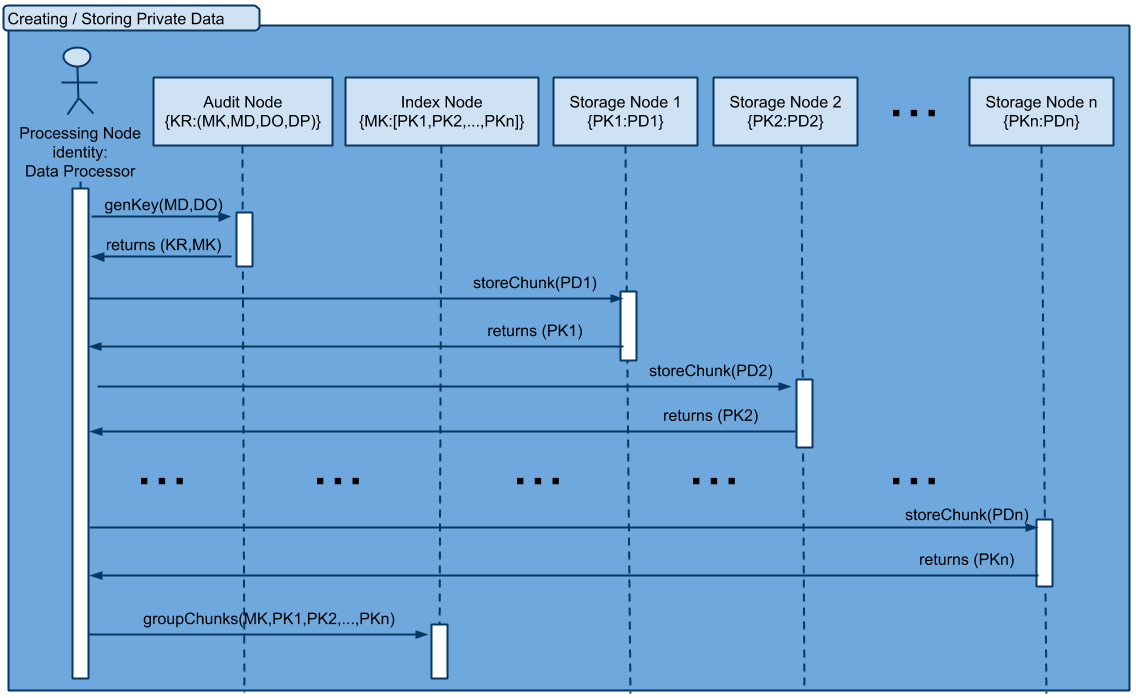}
		\label{fig:storing}
	\end{center}
\end{figure}

The storage of private data is achieved in three phases, illustrated at a higher level in Figure \ref{fig:storing} and detailed in the following schema:\\
\textbf{Phase 1}\\
1. $PN \rightarrow AN: DO, MD$\\
2. $AN: MK := gen()$\\
3. $AN: KR := gen()$\\
4. $AN[KR] := (MK, MD, DO, DP)$\\
5. $AN \rightarrow PN: KR, MK$\\
\textbf{Phase 2}\\
1. $PN: PD_1, PD_2, ..., PD_n := split(PD)$\\
2. $PN:$ chooses randomly n $SNs$\\
3. $PN \rightarrow SN_i: PD_i, i = \overline{1,n}$\\
4. $SN_i: PK_i := gen(), i = \overline{1,n}$\\
5. $SN_i[PK_i] := PD_i, i = \overline{1,n}$\\
6. $SN_i \rightarrow PN: PK_i, i = \overline{1,n}$\\
\textbf{Phase 3}\\
1. $PN \rightarrow IN: MK, PK_1, PK_2, ..., PK_n$\\
2. $IN[MK] := [PK_1, PK_2, ..., PK_n]$\\
3. $PN[alias] := KR$

\squeezeup
\squeezeup
\squeezeup
\squeezeup
\squeezeup
\squeezeup
\squeezeup
\squeezeup

When a processing node needs to store private data, it starts the first phase by sending to an audit node the metadata describing the information it wants to store along with its identity (considered both data owner because it stores its information and data processor because it is the identity which is going to use the associated reference key for data retrieval). The audit node first generates a master key and a key reference, then stores the generated master key, the received metadata, and the received data owner (as both data owner and data processor) under the generated key reference in its key-value database. The audit node completes this phase by sending the generated master key and the key reference to the processing node.

In the second phase, the processing node splits the private information into n undecipherable chunks $PD_1, PD_2, ..., PD_n$ and chooses randomly n storage nodes so that each storage node is responsible for storing a single undecipherable chunk of private data. Each storage node, upon receiving its undecipherable piece of private data, generates a partial key, stores its chunk of information under that key, and sends to the processing node the generated partial key.

In the third phase, the user sends to an index node the master key and its corresponding partial keys. The processing node stores the key reference under an alias because it is needed for subsequent private information retrieval.

\squeezeup
\subsection{Reading / Retrieving Private Data}

\begin{figure}[htbp]
	\begin{center}
		\caption{Retrieving Private Data}\centering
		\includegraphics[width=4.5in]{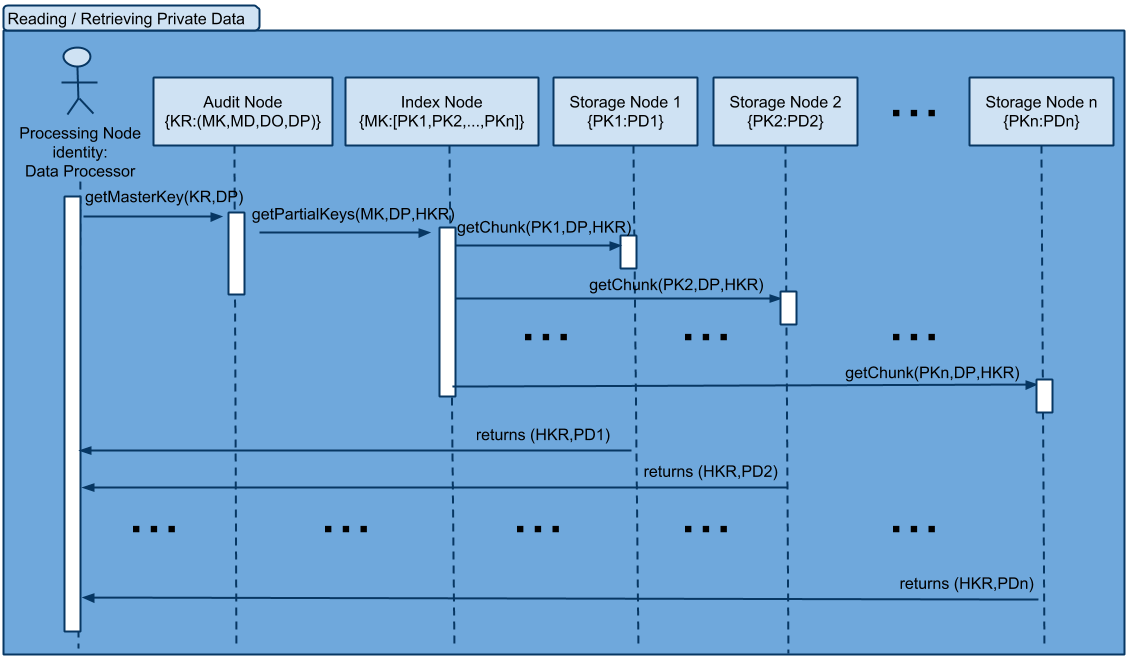}
		\label{fig:retrieving}
	\end{center}
\end{figure}

If a processing node needs to access a private information, it must have a key reference. The way the processing node uses the key reference to retrieve the associated private information can be followed in Figure \ref{fig:retrieving} and is described in detail in the following schema:\\
\textbf{Phase 1}\\
1. $PN \rightarrow AN: DP, KR$\\
\textbf{Phase 2}\\
1. $MK := AN[KR]$\\
2. $HKR := (location(PN), hash(KR))$\\
3. $AN \rightarrow IN: DP, MK, HKR$\\
\textbf{Phase 3}\\
1. $PK_1, PK_2, …, PK_n := IN[MK]$\\
2. $IN \rightarrow SN_i: DP, HKR, PK_i, i = \overline{1,n}$\\
3. $PD_i := SN_i[PK_i], i = \overline{1,n}$\\
4. $SN_i \rightarrow PN: HKR, PD_i, i = \overline{1,n}$\\
5. $PN: PD := recombine(PD_1, PD_2, ..., PD_n)$, where $PD_i, i = \overline{1,n}$ must have the same $HKR$ as $PD$

The key reference might reference either a piece of private information of the processing node or a piece of private information shared to the processing node by another processing node. By sending his key reference to the audit node along with his (processing node's) identity, the processing node completes the first phase.

In phase two, the audit node retrieves the master key corresponding to the received key reference. Next, it computes HKR, which is a hash on the retrieved key reference prefixed with the location of the processing node. Then, the audit node sends the processing node's identity, the retrieved master key, and the computed HKR to the index node. This way, the index node doesn't learn the association between key references and master keys, but at the same time propagates HKR, which is information required by the processing node to identify the request being answered. Note that the processing node might issue multiple data retrieval operations at the same time and, without HKR, the processing node wouldn't know which undecipherable chunks correspond to which pieces of private data he requested at the same time.

In the third phase, the index node retrieves from its database the partial keys corresponding to the master key and sends each partial key along with the processing node's identity and HKR to the corresponding storage nodes. Each storage node retrieves the undecipherable value ($PD_i$) corresponding to the received partial key ($PK_i$) and sends to the processing node the retrieved undecipherable chunk and the HKR. The processing node, upon receiving the undecipherable chunks, groups them by HKR and recombines the grouped components in order to obtain the private piece of information. This information can be processed, but the law prevents the processing node to store it.

\squeezeup
\squeezeup
\squeezeup
\squeezeup
\squeezeup
\squeezeup

Thus HKR's purpose is to serve as an identifier so that a processing node which retrieves multiple private information pieces at the same time can associate the received undecipherable pieces of information to the requested key references.

\squeezeup
\subsection{Updating Private Data}

The first two phases are identical for data retrieving and data updating, but starting with the third step of the third phase, things are performed differently as can be observed in the following schema:
\\\textbf{Phase 1}\\
1. $PN \rightarrow AN: DP, KR$\\
\textbf{Phase 2}\\
1. $MK := AN[KR]$\\
2. $HKR := (location(PN), hash(KR))$\\
3. $AN \rightarrow IN: DP, MK, HKR$\\
\textbf{Phase 3}\\
1. $PK_1, PK_2, ..., PK_n := IN[MK]$\\
2. $IN \rightarrow SN_i: DP, HKR, PK_i, i = \overline{1,n}$\\
3. $SN_i \rightarrow PN: HKR, i = \overline{1,n}$\\
4. $PN: PD_1, PD_2, ..., PD_n := split(PD)$\\
5. $PN \rightarrow SN_i: PD_i, i = \overline{1,n}$\\
6. $SN_i[PK_i] := PD_i, i = \overline{1,n}$

\squeezeup
\squeezeup
\squeezeup
\begin{figure}[htbp]
	\begin{center}
		\caption{Updating Private Data}\centering
		\includegraphics[width=4.5in]{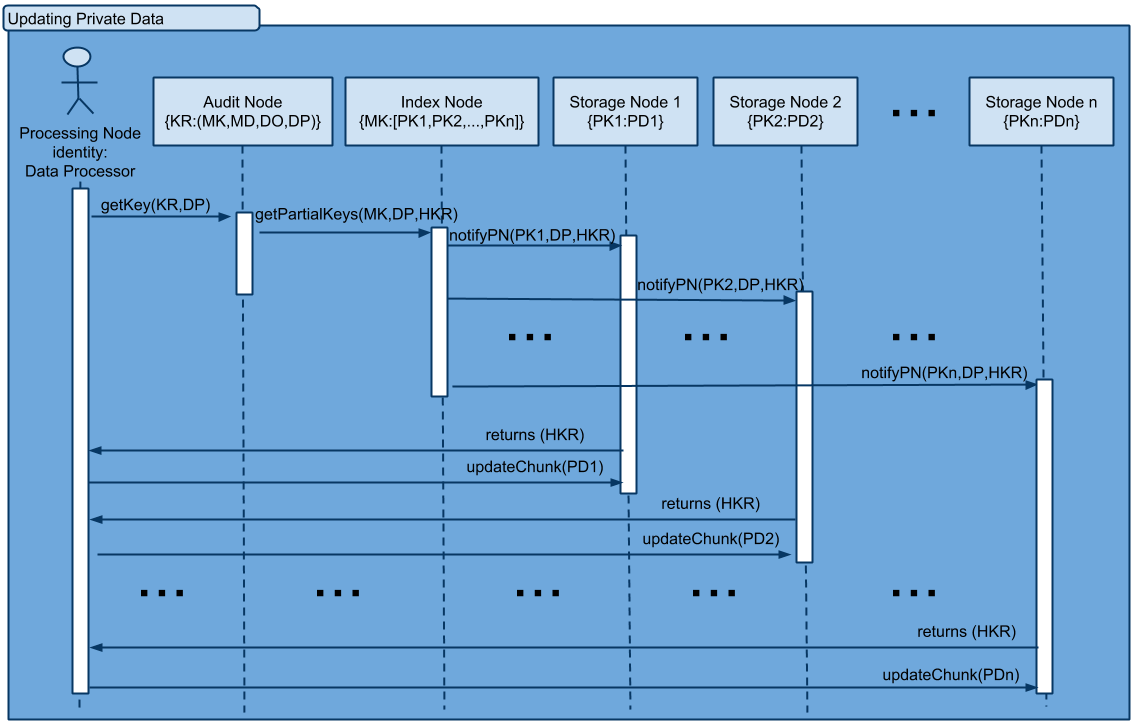}
		\label{fig:updating}
	\end{center}
\end{figure}
\squeezeup
\squeezeup
\squeezeup
The storage nodes, upon receiving the partial keys from the index node, instead of retrieving the undecipherable chunks of private data corresponding to the partial keys and sending them along with HKR to the processing node for recombination as performed by the storing operation, for the updating operation they send the HKR alone to the processing node. Upon receiving the HKR from the storage nodes, the processing node splits the new information in undecipherable chunks and sends one chunk to each storage node which sent the HKR corresponding to this piece of private information. Then, the storage nodes update the values stored under the partial keys in their key-value database in accordance to the newly received undecipherable chunks.

The reason we decided to go with this approach rather than use an invalidation and a store operation is because we want all the existing key references to remain valid and, even more, to point to the updated private data.

The data flow between the nodes which are part of the system during an update operation can be observed in Figure \ref{fig:updating}.

\squeezeup
\subsection{Deleting Private Data}

\begin{figure}[htbp]
	\begin{center}
		\caption{Deleting Private Data}\centering
		\includegraphics[width=3in]{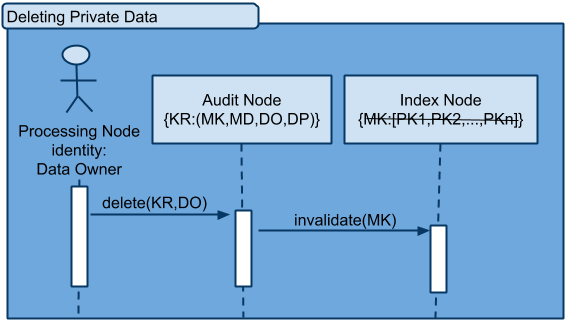}
		\label{fig:deleting}
	\end{center}
\end{figure}

Figure \ref{fig:deleting} illustrates the data flow and the following schema illustrates the actions performed during a delete operation:\\
1. $PN \rightarrow AN: KR, DO$\\
2. $MK := AN[KR]$\\
3. $AN \rightarrow IN: MK$\\
4. $IN:$ invalidate $IN[MK]$

In order to perform a delete operation, a processing node sends to the audit node its identity (which must be the identity of the data owner) and its key reference of the data to be deleted. If the audit node would invalidate the received key reference, this would mean only revoking access to the private data for the data owner, while all the data processors which received access to this private data at some point in time would still be able to access the data. Thus, instead of doing this, the audit node sends the received key reference to the index node for it to invalidate the associated master key. In this way, neither the data owner, nor the data processors will be able to access this piece of private data anymore because all the key references they have for this piece of private data point to the same master key.

\squeezeup
\squeezeup
\squeezeup
\squeezeup
\squeezeup
\squeezeup
\squeezeup
\squeezeup
\squeezeup
\subsection{Sharing Access to Private Data}

\begin{figure}[htbp]
	\begin{center}
		\caption{Sharing Access to Private Data}\centering
		\includegraphics[width=3in]{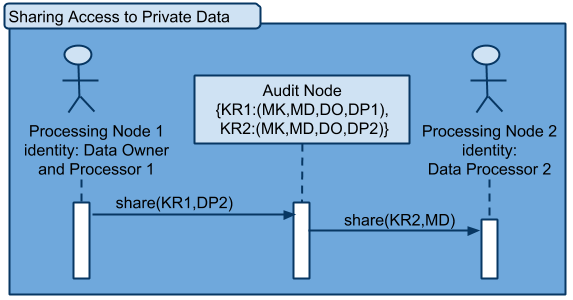}
		\label{fig:sharing}
	\end{center}
\end{figure}

The sharing operation is described in Figure \ref{fig:sharing} and follows the following steps:\\
1. $PN_1 \rightarrow AN: KR_1, DP_2$\\
2. $MK, MD, DO := AN[KR_1]$\\
3. $AN: KR_2 := gen()$\\
4. $AN[KR_2] := (MK, MD, DO, DP_2)$\\
5. $AN \rightarrow PN_2: KR_2, MD$

\squeezeup
\squeezeup
\squeezeup
\squeezeup
\squeezeup
\squeezeup

In order to share a piece of information, a processing node ($PN_1$) must send to an audit node its key reference ($KR_1$) of the private information it wants to share along with the identity of the processing node that is to receive access to the private information ($DP_2$). When this happens, the audit node retrieves the master key ($MK$) corresponding to the received key reference ($KR_1$), generates a new key reference ($KR_2$), and saves the retrieved master key ($MK$), the retrieved metadata ($MD$), the retrieved data owner ($DO$) and the received data processor ($DP_2$) under the newly generated key reference ($KR_2$). Of course, the initial association (between $KR_1$ and $MK$) remains in the database, as well.

Note that every association between a key reference and a master key also has  information regarding the identity of the organization owning the data (Data Owner) and the identity of the organization with which data is shared (Data Processor). If data owner is the same with data processor, then this association is the initial key reference created when the private information was first stored.\\
Next, the audit node sends the newly generated key reference ($KR_2$) along with the received metadata ($MD$) to the processing node which is to receive access ($PN_2$) to the private data. In this way, neither data owner knows the data processor's key reference, nor the data processor knows the data owner's key reference.

\squeezeup
\squeezeup
\subsection{Revoking Access to Private Data}
\squeezeup

\begin{figure}[htbp]
	\begin{center}
		\caption{Revoking Access to Private Data}\centering
		\includegraphics[width=3in]{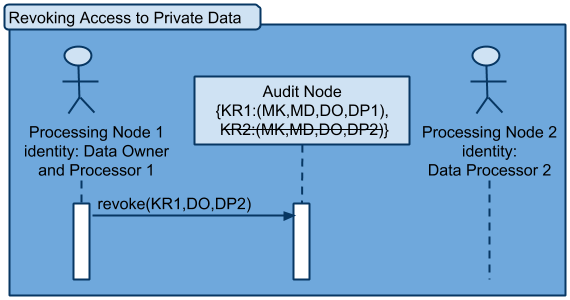}
		\label{fig:revoking}
	\end{center}
\end{figure}

The revocation operation is described in Figure \ref{fig:revoking} and follows the following steps:\\
1. $PN \rightarrow AN: KR_1, DO, DP_2$\\
2. $MK := AN[KR_1]$\\
3. search $KR_2$ which contains $MK, DO, DP_2$ as values\\
4. invalidate $AN[KR_2]$

\squeezeup
\squeezeup
\squeezeup
\squeezeup
\squeezeup
\squeezeup
\squeezeup

The data owner can revoke access to a private information by issuing a revocation request to the audit node. The revocation request contains the identity of the data owner and of the data processor to which access is being revoked as well as the data owner's key reference ($KR_1$). Note that we receive the data owner's key reference, while the revocation needs to be done on data processor's key reference ($KR_2$). This happens because each processing node knows its key reference, but it doesn't know the key references of the data processors which have access to its data. So, the audit node needs to retrieve the master key corresponding to the received key reference ($KR_1$) and search the key reference to be revoked ($KR_2$) knowing that it has associated the retrieved master key and the received data owner and data processor. After learning the value of the reference key to be revoked, the audit node simply invalidates it. Nothing is deleted.

\squeezeup
\squeezeup
\subsection{Copying Private Data}
\squeezeup

By design, any copy operation on the private data should be done only through the sharing operation. Data derived from the private data should be stored in the PDS and assigned to the original data owner.
\squeezeup

\section{System Analysis from the Privacy Perspective}
\squeezeup
\label{sec:system_analysis}
In this section we are going to analyze how powerful each type of node defined in the system is  and how much information they can gather by themselves or by colluding with other types of nodes.

Each storage node has access to only one undecipherable chunk of each private piece of information it stores. Each chunk is saved under a partial key which has no meaning to the storage node. The storage node doesn't know which other storage nodes the other chunks of the same pieces of private information store, nor under which partial keys. Even more, the storage node doesn't know what type of information it stores. It may be a social security number, a password, a name, a birth date, and so on. A single storage node can't attack the system and neither can a collection of colluding storage nodes.

Index nodes store only the associations between master keys and partial keys. So, they know the partial keys whose corresponding undecipherable chunks can be recombined to form a private piece of information, but they don't know the values of the actual chunks, nor the type of information that will be obtained after recombining the chunks. A single index node can't attack the system and neither a collection of colluding index nodes.

Audit nodes have information regarding the meaning of the data, the owner of the data, and the identities with which the data was shared, but they don't have information regarding the way the data was split in chunks (the correspondence between master key and partial keys) and the locations where the data chunks are stored. So, a single audit node or a group of colluding audit nodes can't recombine the private data. However, audit nodes are able to create reference keys at their discretion and share them with legal or illegal organizations.

Processing nodes have access to the private data as they need it for normal operations. Privacy by Design principles are intended to regulate the usage of private data without reducing functionality. The main goal of the PDS is to make it obvious when a company is misusing the private data outside the purpose accepted by the user, but without reducing access to the private data. For example, if an organization collects private data by using PDS, it becomes visible if it is copying or using private data for other purposes than intended.

Only processing nodes and audit nodes know what the pieces of information referenced by key references mean. Encryption is not needed because the attackers see a huge pool of partial undecipherable messages. Traffic can't be used to obtain information because the traffic data is encrypted using TLS and can't be used to deduce information regarding which nodes communicate because of the huge amount of concurrent swarms flying from node to node.

If an index node colludes with all the storage nodes storing chunks of the same piece of private information, then together they can recombine the message, but without knowing its meaning, who owns it and with whom it was shared with, it is of no value to them. In order for the data to be of value, they need to collude also with the audit node, which stores the metadata, the data owner and the data processor of this particular piece of information.
\squeezeup
\section{Conclusions}
\squeezeup
\label{sec:conclusions}
In normal conditions only processing nodes should be able to read plaintext data. All the other node types involved in the PDS should not be capable of accessing private data. In special conditions, audit nodes should be able to read the data as well in order to enable legal access to the private data owned by other data owners for crime prevention or other legal usages. We imagine audit organizations offering public services that are controlled by the law and industry regulations. The level of access to the systems storing this metadata should be similar to the one for financial services. Special legal procedures should be followed when accessing private data outside of the normal flow.

Systems and approaches that are trying to obfuscate and encrypt too much are fighting an impossible fight with the common social interest and are blocking the normal evolution of the technologies in the privacy area. The interests of any citizen are to be protected from unfair usage of his data by the large Internet companies, to have control on who he shares his private data with, to be able to revoke access to his data to anyone at any time.

An Internet based on fully homomorphic encryption would not be what we need because it would create a world in which data can be too easily lost. It would provide a perfect method for criminals and terrorists to hide their data from the public interest. Fighting with dangerous, corrupted governments is important, but PDS is not supposed to have a role in this fight. PDS is a balanced solution which enforces Privacy by Design in code and maintains an equilibrium between public and private interests.
\squeezeup
\section{Future Work}
\squeezeup
\label{sec:future_work}
As future work, we intend to pursue three different paths. First, we will develop a new self-sovereignty authentication technique which uses the advantages provided by the architecture of the system proposed in this paper. Secondly, as Privacy by Design and Privacy by Default (PbD) are being enforced by laws (eg in the General Data Protection Regulation), we intend to propose a Privacy Enhancing Technique (PET) that can ensure these principles directly in code. It is supposed to be a privacy estimation method for systems using the technique proposed in this paper for achieving self-sovereign storage of private data.

Thirdly, we will propose and describe a mechanism for the audit nodes to store the metadata so that it enables the implementation of personal assistants. The metadata will describe the schema of the stored objects (in the form of JSON schema or OWL) and the representation types that could enable type checking when data is shared. It will enable the use of specific Privacy Policies (which will control what entities are allowed to read the information and will contain revocation policies) and Security Policies (which will control what entities are allowed to modify the content of a Master Key). Both privacy and security policies will be enforced by the audit nodes, but the input (rules and policies) will be provided by the Data Owner himself. Giving up to the standard communication promoted by web technologies and moving towards a model of communication verifiable as the one proposed by executable choreographies, we have the opportunity to develop formal verifications methods on how the private data is used.
\squeezeup
\section{Acknowledgments}
\squeezeup
\begin{wrapfigure}{r}{0.10\textwidth}
	\vspace{-27pt}
	\includegraphics[width=\linewidth]{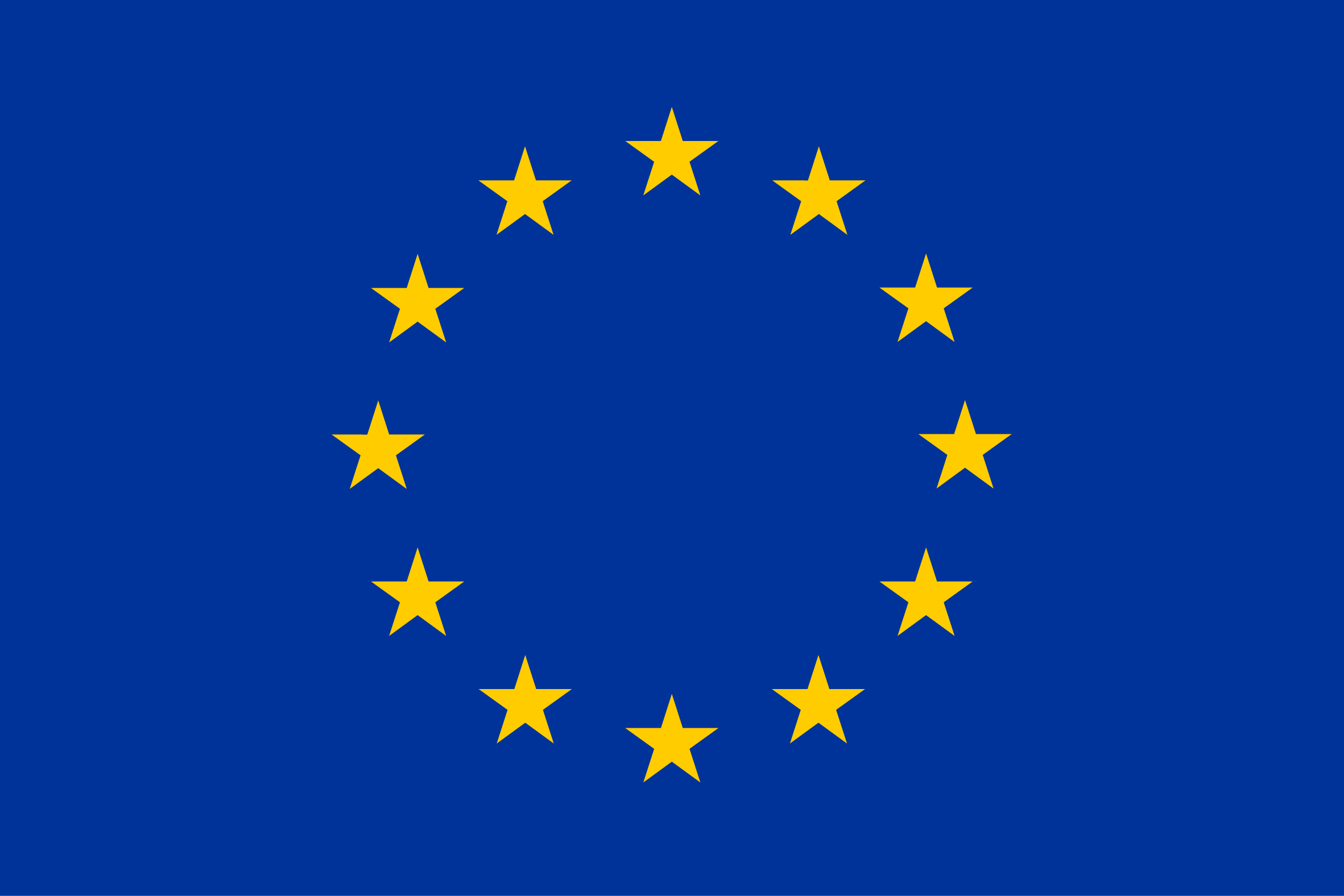}
\end{wrapfigure}
This work is partly funded by the \textbf{European Union's Horizon 2020 Research and Innovation Programme} under grant agreement No 692178.

It is also partially supported by the \textbf{Private Sky Project}, under the \textbf{POC-A1-A1.2.3-G-2015 Programme} (Grant Agreement no. P\_40\_371).

\squeezeup
\bibliography{paper}{}
\bibliographystyle{plain}
\end{document}